\documentclass[a4paper,10pt,twoside]{cpc-hepnp}

\usepackage{multicol}
\usepackage{graphicx}
\usepackage{booktabs}
\usepackage{amssymb,bm,mathrsfs,bbm,amscd}
\usepackage[tbtags]{amsmath}
\usepackage{lastpage}

\begin{document}

\fancyhead[c]{\small  Chinese Physics C~~~Vol. 37, No. 1 (2013)
010201} \fancyfoot[C]{\small 010201-\thepage}

\footnotetext[0]{Received 5 July 2013}

\title{The description of $^{150}$Nd Nucleus by a new alternative scheme
\thanks{Supported by Natural Science Foundation of Liaoning Scientific Committee (2013020091)
and National Natural Science Foundation of China.
 }}

\author{%
 Dai Lian-Rong\email{dailr@lnnu.eu.cn}%
\quad Zhong Wei-Wei%
\quad Cong Mei-Ling%
\quad Wang Li-Xing%
}
\maketitle

\address{%
 Department of Physics, Liaoning Normal University, Dalian, 116029, China}

\begin{abstract}
A new scheme was recently proposed  in which the usual $SU(3)$ quadrupole-quadrupole
interaction was replaced by an $O(6)$ cubic interaction in
the Interacting Boson Model, and also successfully applied to the description of $^{152}$Sm for the N = 90 rare earth isotones
with X(5) symmetry.  By using this new scheme, in the
present work, we further  explore the properties of another candidate of $^{150}$Nd for the $N=90$  with X(5) symmetry.  The low-lying energy levels and E2 transition rates are calculated and compared with the experimental data. The results show  that the new scheme can also reasonably describe the experimental low-lying spectrum and the intraband and the interband E2 transitions for $^{150}$Nd.
However, for the low-lying spectrum, the $O(6)$ cubic interaction seems  better in describing the energy levels, especially  in higher excited states and $\gamma$ band,
yet the  $0^{+}_{2}$ level within the $\beta$ band is lower than the corresponding experimental value and, the $U(5)$-$SU(3)$ scheme seems better to describe the
low-lying levels of $\beta$ band;  and for the B(E2) transition, for the intraband transitions within the ground band and some interband transitions between the $\beta$ band and the ground band,  the results from  $O(6)$ cubic interaction are better than those from $SU(3)$ quadrupole-quadrupole interaction, yet of which seems better to describe the intraband E2 transitions within $\beta$ band.  The present work is very meaningful in helping us to deeply understand the
new characteristics of symmetry  by the higher order $O(6)$ cubic interaction.
\end{abstract}



\footnotetext[0]{\hspace*{-3mm}\raisebox{0.3ex}{$\scriptstyle\copyright$}2013
Chinese Physical Society and the Institute of High Energy Physics
of the Chinese Academy of Sciences and the Institute
of Modern Physics of the Chinese Academy of Sciences and IOP Publishing Ltd}%

\begin{multicols}{2}

\section{Introduction}
{The nature of shape phase transitions in finite many-body systems is
a fundamental issue and has been the subject of many investigations~\cite{sa,rowe05, tim, for}.
Transitional nuclei experienced renewed interest over the last decade and  were extensively studied in the Interacting
Boson Model (IBM)~\cite{lab1,lab2,lab3,lab4,lab5}. It is now widely accepted that the three limiting cases~\cite{af1,af2,af3}
of the theory correspond to three different geometric shapes of nuclei,
referred to as spherical (vibrational limit with U(5) symmetry), axially
deformed (rotational limit with SU(3) symmetry), and $\gamma$-soft
(triaxial with O(6) symmetry), respectively.
The full range of the model can be parameterized in terms of
the Casten triangle~\cite{castri}. Interesting phenomena occur
when a system falls between two limits of the theory, in which case a
quantum phase transition occurs at a critical point~\cite{die1,die2,feng}
where the dominance of one of the symmetries yields to the dominance of
the other. And indeed, the $U(5)$-$SU(3)$ transitional description for
Nd, Sm, Gd, and Dy nuclei was first reported
in Ref.~\cite{sch}. Later on, the $X(5)$ symmetry at the critical
point of this transition was studied in Ref.~\cite{iac1}, in which
Iachello proposed analytical solutions of a Bohr Hamiltonian
in the situation appropriate for the description of nuclei near the critical
point of the spherical to axially deformed shape-phase transition.
Casten and Zamfir showed in Ref.~\cite{cas2}
that $^{152}$Sm and other $N=90$ isotones demonstrate these characteristics.}
Transitional patterns from the spherical, U(5), to the axially deformed,
SU(3), limit of the IBM with a schematic Hamiltonian  were studied
in Refs.~\cite{fp,ros05,fos}; in particular, the transitional behavior of some physically
relevant quantities across the entire span of the $U(5)$-$SU(3)$
transitional region were explored.

In fact another new idea was first suggested by van Isacker~\cite{isa}, in which
the quantum phase transitional behavior for an alternative characterization of the
spherical to axially deformed shape-phase
transition was analyzed. In Ref.~\cite{isa}, the well-known $SU(3)$
quadrupole-quadrupole interaction in the
schematic Hamiltonian ($U(5)$-$SU(3)$ scheme)
is replaced  by  the $O(6)$ $[\hat{Q}{(0)}\times \hat{Q}{(0)}\times \hat{Q}{(0)}]^{0}$
cubic interaction, where  $\hat{Q}_{\mu}{(0)}=s^{\dagger}\tilde{d}_{\mu}+
d^{\dagger}_{\mu}s$ is generator of the $O(6)$ algebra.
This idea was  then further confirmed
by Thiamova and Cejnar~\cite{tc}. From these investigations one understands that
similar results may be realized when the quadratic scalar formed with the $U(5)$-$SU(3)$
quadrupole operator is replaced by the cubic scalar formed with the quadrupole interaction
of the $O(6)$ limit.  Inspired by this new idea, in our very recent work~\cite{Dai:2012sc}, a systematic investigation of this
alternative scheme was dynamically investigated, in which the transitional behaviors of the low-lying energy levels, eigenstates,
isomer shifts, E2 transition rates, and expectation values of shape variables
across the entire transitional region  are all examined. A comparison with outcomes of the
usual $U(5)$-$SU(3)$ transitional description
shows that the spherical to axially deformed shape-phase transition can also be described
within this alternative context, especially near the critical point. The results show that
the transition with the $O(6)$ cubic interaction (UQ scheme) is considerably smoother than the $U(5)$-$SU(3)$
case, but with the nuclear shape less well defined,
even in the axially deformed limit. It is also shown that the new scheme can also reasonably describe the low-lying spectrum
and  E2 transition rates of $^{152}$Sm for $N=90$, and the new scheme seems
better than the usual $U(5)$-$SU(3)$ scheme in describing the properties of $^{152}$Sm
with the $X(5)$ symmetry.

The purpose of the present study is to further investigate another candidate of N=90 rare earth isotones with X(5) symmetry.
The low-lying energy levels and the intraband and the interband E2 transitions relating ground, $\beta$, and $\gamma$ bands for $^{150}$Nd
will be calculated within both the original $U(5)$-$SU(3)$ scheme and UQ scheme, and compared with the experimental data.
We will show that the typical quantities in the new UQ scheme, including the low-lying spectrum and E2 transitions for $^{150}$Nd, indeed display almost the same transitional patterns as  found in the  original $U(5)$-$SU(3)$ scheme. The analysis  will further confirm that the $O(6)$ cubic interaction
can play a role similar to that of the $SU(3)$ quadrupole-quadrupole interaction.

\section{Hamiltonian}

It is well known that the consistent-$Q$ Hamiltonian~\cite{Warner1983}
can be used to describe the most general situation in the IBM, which is given by

\begin{equation} \label{Hamiltonian-IBM}
\hat{H}_{\rm Q}=c_{1} \left[ x\hat{n}_{d} +
\frac{x-1}{f_{1}(N)}\hat{Q}({\chi}) \cdot \hat{Q}({\chi}) \right] \, ,
\end{equation}
where $N$ is the total number of bosons, $\widehat{n}_{d}$
is the number operator  for counting $d$-bosons,
$\hat{Q}_{\mu}({\chi}) = (d^{\dag} s + s^{\dag} \tilde{d})^{(2)}_{\mu} +
\chi (d^{\dag} \tilde{d})^{(2)}_{\mu}$,  $x$ and $\chi$ are control
parameters, $f_{1}(N)$ is a linear function of $N$, and $c_{1}$ is the scaling factor.
Therefore, three limit situations of the model are given by
$x=1$ for the $U(5)$, $x=0$ and $\chi=0$ for the $O(6)$, and
$x=0$ and $\chi=-\frac{\sqrt{7}}{2}$ for the $SU(3)$, respectively.

\subsection{ For $U(5)$-$SU(3)$ scheme}

The original $U(5)$-$SU(3)$ scheme~\cite{fp}  is the special case of the consistent-$Q$ formalism (\ref{Hamiltonian-IBM}) with $\chi=-\frac{\sqrt{7}}{2}$, which is
suitable to describe the spherical to axially deformed shape phase transition,
thus the Hamiltonian is given by
\begin{eqnarray}
\hat{H_1}=c_1\left[x\hat{n}_{d}+\frac{(x-1)}{f_1(N)}\hat{Q}(-{\sqrt{7}/{2}})\cdot \hat{Q}(-{\sqrt{7}/{2}})\right],\label{h1}
\end{eqnarray}
where $f_{1}(N)$ is a linear function of $N$, and as shown in Refs.~\cite{iac1,fp,wer},
the critical point $x_c$ will be different for different choices
of the function $f_1(N)$, we adopt $f_1(N)=4N$ as used
in \cite{iac1,fp,wer}.  And parameter $c_{1}>0$ is the scaling factor, $x$ is the control parameter of the $U(5)$-$SU(3)$ scheme with
$0\leq x\leq 1$.

The B(E2) transition is a very sensitive signature of the structure. The B(E2) reduced  transition probability is defined by
\begin{equation}
B(E2;L_i \rightarrow L_f) =\frac{|\langle L_f \| T(E2) \| L_i\rangle|^2 }{2L_i+1}
\end{equation}
here the $T(E2)$ operator in the $U(5)$-$SU(3)$ scheme is chosen to be
the same as that used in Ref.~\cite{fp} with
\begin{equation}\label{T1}
T_{\mu}(E2)=q_{1}\hat{Q}_{\mu}(-{\sqrt{7}/{2}}),
\end{equation}
where
\begin{equation}
\hat{Q}_{\mu}(-{\sqrt{7}/{2}})=(s^{\dagger} \tilde{d}_{\mu}+d^{\dagger}_{\mu} {s})-\sqrt{7}/2
(d^{\dagger} \tilde{d})^{(2)}_{\mu}
\end{equation}
 is the SU(3) generator, and
$q_{1}$ is the effective charge.

\subsection{ For UQ scheme}

Alternatively, in  the UQ scheme,  the suitable Hamiltonian to describe the same shape phase transition in this region is given by
\begin{eqnarray}
\hat{H_2}=c_2\left[y\hat{n}_{d}+\frac{(1-y)}{f_2(N)}[Q{(0)}\times
Q{(0)}\times Q{(0)}]^{0}\right]. \label{h2}
\end{eqnarray}
where parameter $c_{2}>0$ is the scaling factor, $y$ is the control parameter of the UQ  scheme with
$0\leq y\leq 1$,  $f_{2}(N)$ is a quadratic function of $N$, the critical point $y_c$ will be different for different choices of the function $f_2(N)$, we adopt $f_2(N)=0.8N^2$ as used
in Ref.~\cite{Dai:2012sc} which puts the critical point $y_{\rm c}$
of the UQ scheme close to $x_{\rm c}$ of the $U(5)$-$SU(3)$ scheme.

The $T(E2)$  operator in the UQ scheme is chosen to be
\begin{equation}\label{T2}
T_{\mu}(E2)=q_{2}\hat{Q}_{\mu}(0),
\end{equation}
where
\begin{equation}
\hat{Q}_{\mu}{(0)}=s^{\dagger}\tilde{d}_{\mu}+
d^{\dagger}_{\mu}s
\end{equation}
is the $O(6)$ quadrupole operator and $q_{2}$ is the effective charge.

\subsection{How to solve equation}

In order to diagonalize Hamiltonians (\ref{h1}) and
(\ref{h2}), we expand the corresponding
eigenstates in terms of the $U(6)\supset
SU(3)\supset SO(3)$ basis vectors $|N(\lambda\mu)KL\rangle$ as
\begin{eqnarray}
|NL_{\xi}\rangle=\sum_{(\lambda\mu)K}C^{L_{\xi}}_{(\lambda\mu)K}|N(\lambda\mu)KL\rangle,
\label{wf}
\end{eqnarray}
where $\xi$ is used to denote the $\xi$-th level with angular momentum
quantum number $L$, and the
$C^{L_{\xi}}_{(\lambda\mu)K}$ are expansion coefficients. Since the
total number of bosons $N$ is fixed for a given nucleus, the
eigenstates given in Eq.~ (\ref{wf}) is also denoted
$|L_{\xi};z\rangle$ with $z=x$ for the $U(5)$-$SU(3)$ scheme or $z=y$ for the UQ scheme
in the following,  {where} the value of the
control parameter $z$ is explicitly shown.  In our calculations, the
orthonormalization process~\cite{jpd,aki} with respect to the additional
quantum number $K$ needed to label the basis vectors
of $SU(3)\supset SO(3)$ and
the phase convention for the  $U(6)\supset SU(3)$ basis vectors
proposed in Ref.~ \cite{ros} are adopted. By using analytic expressions
for  $U(6)\supset SU(3)$ reduced matrix elements of the $d$-boson
creation or annihilation operator~\cite{ros} and an algorithm \cite{jpd,aki}
for generating the  $SU(3)\supset SO(3)$ Wigner coefficients, the
eigenequation that simultaneously determines the eigenenergy and the
corresponding set of the expansion coefficients
${C^{L_{\xi}}_{(\lambda\mu)K}}$ can be established, {with
results that can then} be used to calculate physical quantities
in both schemes.

\section{Results and Discussion}

In order to demonstrate that the UQ scheme indeed serves
 as an alternative description of the spherical to axially deformed
 phase transition, recently we have systematically investigated the possible $X(5)$
 candidate  of $^{152}$Sm for $N=90$,
 and shown that the new scheme can indeed reasonably describe the low-lying spectrum of $^{152}$Sm and
 E2 transition rates with the $X(5)$ critical point symmetry \cite{Dai:2012sc} .

 Now in this subsection, another possible $X(5)$
symmetry candidate, $^{150}$Nd nucleus,  will be further investigated by the new UQ scheme, of which the results are compared with
 the experimental data~\cite{Nd150,A150,k,w3} and those obtained from the $U(5)$-$SU(3)$ scheme.
 In the following, we present the calculated
results in both $U(5)$-$SU(3)$ and UQ schemes, including some low-lying
energy levels, the intraband and the interband E2 transitions relating ground, $\beta$, and $\gamma$ bands.
Since the proton number is $60$  and the neutron number is 90, so the total number of bosons for $^{150}$Nd is $N=9$. In the calculation,
the taken value of $x$ for the $U(5)$-$SU(3)$ scheme and $y$ for the UQ scheme  will be decided by the
experimental low-lying levels of $^{150}$Nd.

Firstly, the low-lying levels will be directly obtained for $^{150}$Nd by solving the Hamiltonian (\ref{h1}) for $U(5)$-$SU(3)$ scheme or
Hamiltonian (\ref{h2}) for UQ scheme.

We must choose the parameters. In order to achieve global
quality of fits to low-lying spectrum, the parameters $c_1$ and $x$
for the $U(5)$-$SU(3)$ scheme, or parameters $c_2$ and $y$
for the UQ scheme,  are chosen when the mean-square deviation for excitation energies
\begin{eqnarray}
\sigma({\rm E})=\sqrt{\sum_{i}^{{\cal N}_{1}}|E_{exp}^{i}-E_{th}^{i}|^{2}/{\cal N}_1}.
\end{eqnarray}
reaches the corresponding minimum, where $E^{i}_{\rm th}$, and  $E^{i}_{\exp}$ are energy of the $i$-th level
calculated, and that of the corresponding experimental value,
respectively, and ${\cal N}_{1}$ is the total number of levels fitted.

In Table~\ref{tab1}, we show the  calculated results from both the $U(5)$-$SU(3)$ scheme and the UQ scheme of
low-lying excitation energies  $E(L_{i}^{+})$ (in keV) normalized to the $2_{1}^{+}$ state, and the corresponding mean-square deviation $\sigma(E)$ for excitation energies.
It is shown that $\sigma(E)$ deviation  is 80 keV and 72 keV for $U(5)$-$SU(3)$ scheme and the UQ scheme, respectively, indicating that the overall fitting of low-lying excitation energies from UQ scheme is better than those of $U(5)$-$SU(3)$ scheme.  This can also be clearly seen in Fig.~\ref{fig1}, in which the low-lying levels from the ground band, $\beta$ and $\gamma$ bands are drawn, the experimental data and the calculated results of $U(5)$-$SU(3)$ scheme and UQ scheme are represented, respectively. We can see that both the $U(5)$-$SU(3)$ scheme and the UQ scheme can reasonably describe the low-lying levels of $^{150}$Nd. However, the UQ scheme seems  better in describing the low-lying spectrum, especially in higher excited levels and $\gamma$ band, yet the
$0^{+}_{2}$ level is lower than the corresponding experimental value, indicating that the $U(5)$-$SU(3)$ scheme seems better to describe the levels of $\beta$ band.

\begin{center}
\tabcaption{ \label{tab1} The energies levels (in keV) for $^{150}$Nd.}
\footnotesize
\begin{tabular*}{80mm}{c@{\extracolsep{\fill}}ccc}
\toprule  &    &  $E(L_{i}^{+})$ &  \\\cline{2-4}
$L_{i}^{+}$ & Exp~\cite{k,w3}  & $U(5)$-$SU(3)$ & UQ \\
\hline
$2_{1}^{+}$\hphantom{00} &\hphantom{0} 130.2 & \hphantom{0}130.2 & \hphantom{0} 130.2\\
$4_{1}^{+}$\hphantom{00} &\hphantom{0} 381.1 & \hphantom{0} 382.0 & \hphantom{0} 381.7\\
$6_{1}^{+}$\hphantom{00} &\hphantom{0} 720.4 &\hphantom{0} 741.1 & \hphantom{0} 736.4 \\
$8_{1}^{+}$\hphantom{00} &1129.6 & 1198.0 & 1171.2\\
$10_{1}^{+}$\hphantom{00}&1598.5 & 1746.4 & 1659.4\\
$0_{2}^{+}$\hphantom{00} &\hphantom{0} 675.9 & \hphantom{0} 610.7 & \hphantom{0} 460.1  \\
$2_{2}^{+}$\hphantom{00} &\hphantom{0}850.8 & \hphantom{0} 879.9 & \hphantom{0} 790.5  \\
$4_{2}^{+}$\hphantom{00} &1137.8 & \hphantom{0} 1245.0 & 1165.2 \\
$2_{3}^{+}$\hphantom{00} &1062.1 & \hphantom{0} 1090.6 & 1058.7 \\
$4_{3}^{+}$\hphantom{00} &1350.5 & \hphantom{0} 1510.2 & 1379.1 \\
$\sigma(E)$\hphantom{00} &  & \hphantom{0} 80.0 & 72.0 \\
\bottomrule
\end{tabular*}
\end{center}

\begin{center}
\includegraphics[width=9.5cm]{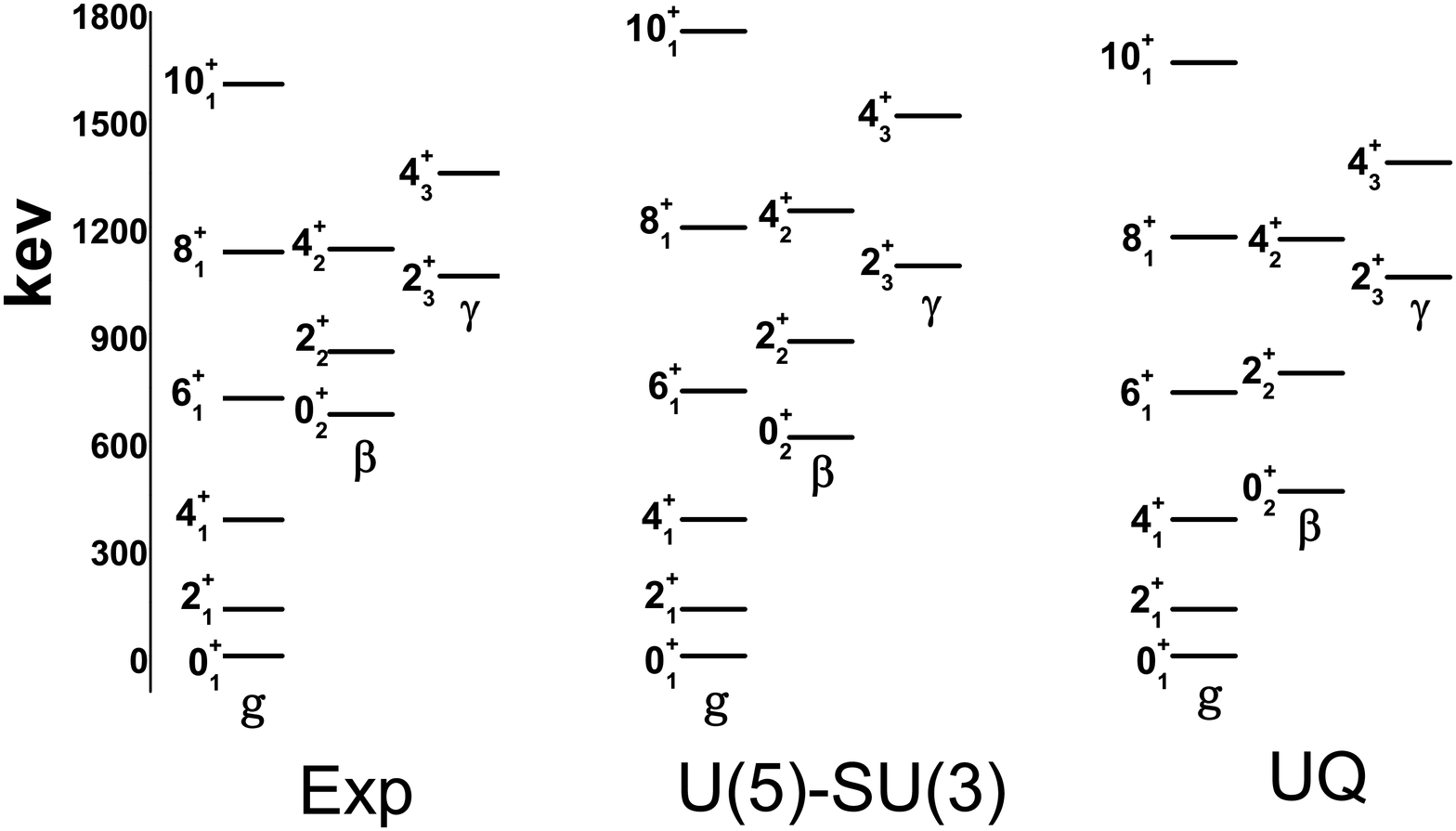}
\figcaption{\label{fig1}
The energies levels  for $^{150}$Nd.}
\end{center}


Secondly, B(E2) transition, as another quantity which acts as a sensitive signature of the structure, will be further studied.
In fact in Ref.~\cite{k} the experimental reduced transition probabilities in $^{150}$Nd were obtained and  compared to the predictions of the critical point symmetry X(5) of the phase shape transition that occurs for the $N=90$ rare earth isotones,  and the very good agreement was observed, revealing this as the case for the realization of the X(5) symmetry.

Here in the present work, we will use  both the $U(5)$-$SU(3)$ scheme and the UQ scheme to study the B(E2) transition. The values of the intraband and the interband E2 transitions relating ground, $\beta$, and $\gamma$ bands will be calculated for $^{150}$Nd using the above constructed E2 transition operators from Eq.~(\ref{T1}) and
Eq.~(\ref{T2}) for  $U(5)$-$SU(3)$ scheme and the UQ scheme, respectively.  In Table~\ref{tab2},
we list the E2 transitions of experimental data~\cite{Nd150,A150,k,w3} and calculated results from both the $U(5)$-$SU(3)$ scheme and the UQ scheme for $^{150}$Nd, in which the B(E2) values~(in W.u.) are normalized to the $2_{1}^{+}\rightarrow 0_{1}^{+}$ transition. The corresponding mean-square deviation
\begin{eqnarray}
\sigma({\rm BE2})=\sqrt{\sum_{i}^{{\cal N}_{2}}|B(E2)_{exp}^{i}-B(E2)_{th}^{i}|^{2}/{\cal N}_2}.
\end{eqnarray}
where ${\cal N}_2$ is the total number of transitions  calculated in the present work, and the deviation is 20.1 W.u. for $U(5)$-$SU(3)$  scheme and 25.6 W.u. for the UQ scheme. It seems  that
both schemes can reasonably describe the transition rates of $^{150}$Nd, however the UQ scheme is  better in describing the intraband transitions  within the ground band, and some interband transitions between the $\beta$ band and the ground band.
Yet the $U(5)$-$SU(3)$ scheme seems  better in describing the intraband E2 transitions within the $\beta$ band.

 In Fig.\ref{fig2}, we  shows some typical  B(E2) values for $^{150}$Nd, the experimental data and calculated results from both the $U(5)$-$SU(3)$ scheme and the UQ scheme are presented, where (a), (b), (c), and (d) represent $B(E2;4_{1}^{+}\rightarrow 2_{1}^{+})$, $B(E2;10_{1}^{+}\rightarrow 8_{1}^{+})$, $B(E2;2_{2}^{+}\rightarrow 0_{1}^{+})$,  and $B(E2;4_{2}^{+}\rightarrow 2_{1}^{+})$, respectively.
 It is clearly shown in Fig. \ref{fig2} that the intraband transitions within the ground band, $B(E2;4_{1}^{+}\rightarrow 2_{1}^{+})$ and $B(E2;10_{1}^{+}\rightarrow 8_{1}^{+})$, the results from  UQ scheme are a little better than those from $U(5)$-$SU(3)$ scheme. While the interband transitions between the $\beta$ band and the ground band, $B(E2;2_{2}^{+}\rightarrow 0_{1}^{+})$ and $B(E2;4_{2}^{+}\rightarrow 2_{1}^{+})$,  especially  the $B(E2;2_{2}^{+}\rightarrow 0_{1}^{+})$ transition  calculated from the two schemes are different.  From Table~\ref{tab2}, the experimental data for $B(E2;2_{2}^{+}\rightarrow 0_{1}^{+})$ transition is 0.7 W.u., however the
  calculated result from  $U(5)$-$SU(3)$ scheme is 0.03 W.u., one order of magnitude smaller than the experimental data. It is improved quite a lot  with
  the  result of  0.16 W.u. in the UQ scheme.  Because the value of $B(E2;2_{2}^{+}\rightarrow 0_{1}^{+})$ from the $U(5)$-$SU(3)$ scheme is too small in comparison with the experimental data,  it seems that the result from UQ scheme is much better than those from  $U(5)$-$SU(3)$ scheme in describing this intraband transition.

\begin{center}
\tabcaption{ \label{tab2} The B(E2) values~(in W.u.) normalized to the $2_{1}^{+}\rightarrow 0_{1}^{+}$ transition  for $^{150}$Nd.}
\footnotesize
\begin{tabular*}{80mm}{c@{\extracolsep{\fill}}ccc}
\toprule  &    &  B(E2) &  \\\cline{2-4}
$L_{i}^{+}\rightarrow L_{f}^{+}$& Exp  & $U(5)$-$SU(3)$ & UQ \\
\hline
$2_{1}^{+}\rightarrow 0_{1}^{+}$\hphantom{00} &\hphantom{0} 116.0 & \hphantom{0}116.0& \hphantom{0} 116.0\\
$4_{1}^{+}\rightarrow 2_{1}^{+}$\hphantom{00} &\hphantom{0} 180.7 & \hphantom{0} 172.8 & \hphantom{0} 173.7\\
$6_{1}^{+}\rightarrow 4_{1}^{+}$\hphantom{00} &\hphantom{0} 206.0 &\hphantom{0} 190.8 & \hphantom{0} 192.0 \\
$8_{1}^{+}\rightarrow 6_{1}^{+}$\hphantom{00} & 216.0 & 192.6 & 195.0\\
$10_{1}^{+}\rightarrow 8_{1}^{+}$\hphantom{00}& 201.0 & 182.8 & 188.2\\
$2_{2}^{+} \rightarrow 0_{2}^{+}$\hphantom{00} &\hphantom{0} 113.9 & \hphantom{0} 71.0 & \hphantom{0}56.6  \\
$4_{2}^{+} \rightarrow 2_{2}^{+}$\hphantom{00} &\hphantom{0} 170.2 & \hphantom{0} 108.5 & \hphantom{0} 89.7  \\
$0_{2}^{+} \rightarrow 2_{1}^{+}$\hphantom{00} & 39.1 & \hphantom{0} 44.3& 32.1 \\
$2_{2}^{+} \rightarrow 0_{1}^{+}$\hphantom{00} & 0.7 & \hphantom{0} 0.03 & 0.16 \\
$2_{2}^{+} \rightarrow 2_{1}^{+}$\hphantom{00} & 10.0 & \hphantom{0} 8.6 &7.82 \\
$2_{2}^{+} \rightarrow 4_{1}^{+}$\hphantom{00} & 19.0 & \hphantom{0} 15.0 & 8.06 \\
$4_{2}^{+} \rightarrow 2_{1}^{+}$\hphantom{00} & 0.015 & \hphantom{0}0.067 & 0.044 \\
$4_{2}^{+} \rightarrow 4_{1}^{+}$\hphantom{00} & 7.015 & \hphantom{0} 6.7 & 10.61 \\
$4_{2}^{+} \rightarrow 6_{1}^{+}$\hphantom{00} & 9.2 & \hphantom{0} 9.86 & 2.51 \\
$2_{3}^{+} \rightarrow 0_{1}^{+}$\hphantom{00} & 3.0 & \hphantom{0} 2.23& 12.15 \\
$2_{3}^{+} \rightarrow 2_{1}^{+}$\hphantom{00} & $> 2.9$ & \hphantom{0} 1.77 & 14.48 \\
$2_{3}^{+} \rightarrow 4_{1}^{+}$\hphantom{00} & 1.7 & \hphantom{0} 4.18 & 4.20 \\
$\sigma(BE2)$\hphantom{00} &  & \hphantom{0} 20.1 & 25.5 \\
\bottomrule
\end{tabular*}
\end{center}

\section{Conclusion}
In the present work, the quantum phase transitional behavior of an alternative
characterization of the spherical to axially deformed shape-phase transition in
IBM is further explored, in which the usual $SU(3)$ quadrupole-quadrupole
interaction is replaced by an $O(6)$ cubic interaction.  We apply this alternative scheme to
further investigate another candidate $^{150}$Nd of N=90 with X(5) symmetry.  The low-lying energy levels and E2 transitions are calculated and compared with the experimental data, and the results show  that the new scheme can also reasonably describe the experimental low-lying spectrum and E2 transitions for $^{150}$Nd.
  Thus the UQ scheme can display almost the same transitional patterns
as  found in the  original $U(5)$-$SU(3)$ scheme. Our analysis confirm that the $O(6)$ cubic interaction
can indeed play a role similar to that of the $SU(3)$ quadrupole-quadrupole interaction.

However, for the low-lying spectrum, the UQ scheme seems  better in describing the energy levels, especially in higher excited levels and $\gamma$ band, yet the $U(5)$-$SU(3)$ scheme seems better to describe the $\beta$ band;  and for the B(E2) transition, the results of the intraband transitions within the ground band from UQ scheme seems  better than  those from $U(5)$-$SU(3)$ scheme,  the UQ scheme can also improve some interband transitions between the $\beta$ band and the ground band. Yet the $U(5)$-$SU(3)$ scheme seems better to describe the intraband E2 transitions within  $\beta$ band.

And indeed, as shown in the application of the $U(5)$-$SU(3)$ and UQ schemes to
the critical point symmetry candidate, $^{150}$Nd, the overall fitting quality of
the UQ scheme is almost similar to that of the $U(5)$-$SU(3)$ scheme.
As has been stated in \cite{isa}, whether the $O(6)$  cubic interaction
in place of the usual $SU(3)$ quadrupole-quadrupole interaction
is required in the deformed limit is not clear at the moment since comprehensive
phenomenological studies of this question are still lacking, hence this is our aim to
provide  more theoretical studies  by the $O(6)$ cubic interaction in comparison with the
$SU(3)$ quadrupole-quadrupole interaction and  experimental data.  It is important that more candidates with
X(5) symmetry should be further explored along this idea to help us deeply understand the
new characteristics of symmetry by the higher order $O(6)$ cubic interaction.

\begin{center}
\includegraphics[width=10cm]{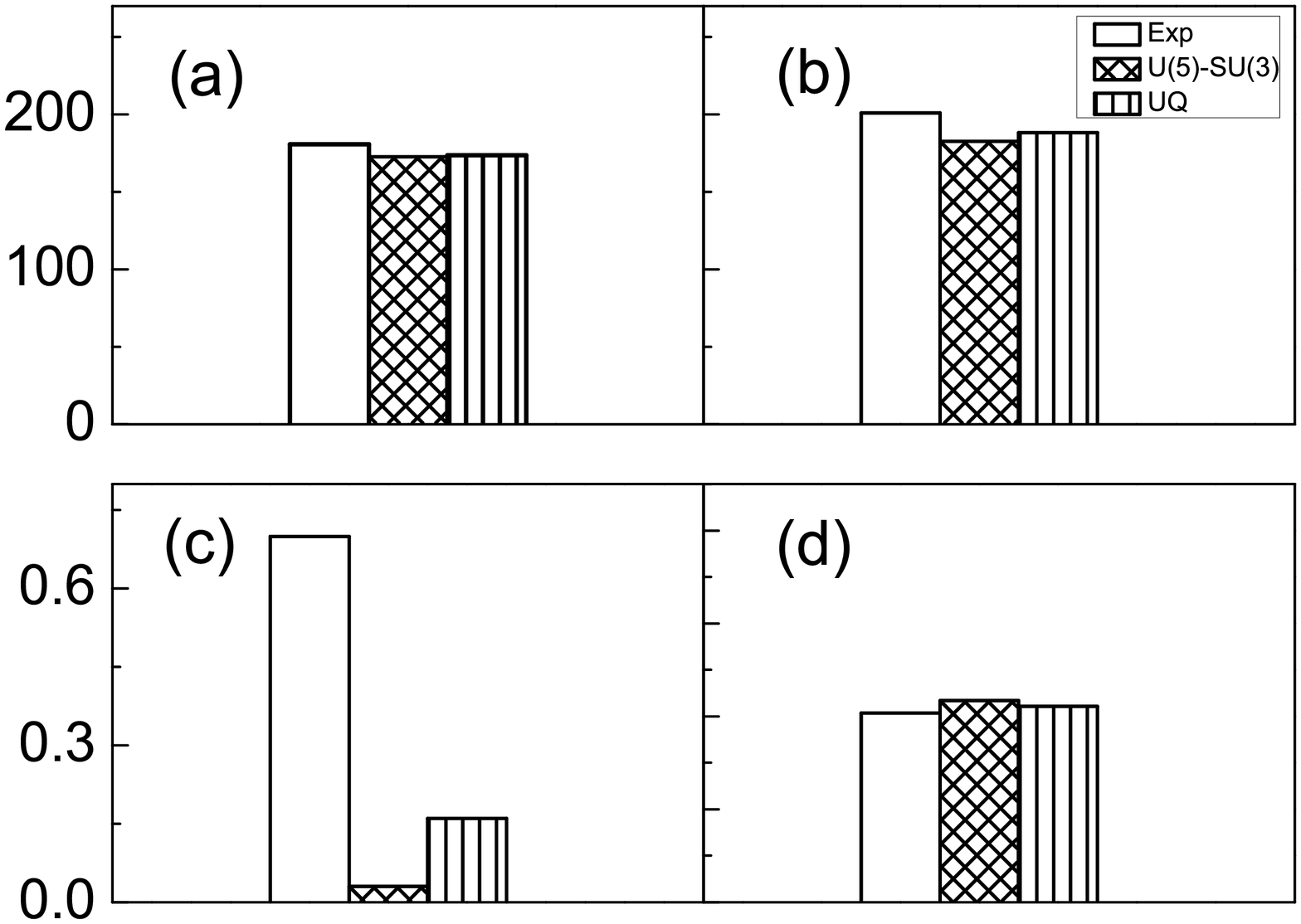}
\figcaption{\label{fig2}
B(E2) values calculated from the $U(5)$-$SU(3)$ scheme and the UQ scheme
  and the corresponding experimental data of $^{150}$Nd, and where
(a) $B(E2;4_{1}^{+}\rightarrow 2_{1}^{+})$, (b) $B(E2;10_{1}^{+}\rightarrow 8_{1}^{+})$, (c) $B(E2;2_{2}^{+}\rightarrow 0_{1}^{+})$,
(d) $B(E2;4_{2}^{+}\rightarrow 2_{1}^{+})$.  }
\end{center}

\end{multicols}


\vspace{-1mm}
\centerline{\rule{80mm}{0.1pt}}
\vspace{2mm}

\begin{multicols}{2}

\end{multicols}

\clearpage

\end{document}